\documentclass[12pt]{iopart}
\usepackage{graphicx,subfig,iopams,wasysym}
\usepackage{cite,url}

\begin{document}

\title[Vibrational relaxation and the non-equilibrium vibrational decomposition of CO$_2$]{Vibrational relaxation and triggering of the non-equilibrium vibrational decomposition of CO$_2$ in gas discharges}
\author{Vladislav Kotov}
\address{Forschungszentrum J\"ulich GmbH, Institut f\"ur Energie- und Klimaforschung - Plasmaphysik (IEK-4),
Partner of the Trilateral Euregio Cluster (TEC), 52425 J\"ulich, Germany}
\ead{v.kotov@fz-juelich.de}

\date{\today}

\begin{abstract}
Non-equilibrium vibrational dissociation of CO$_2$ 
at low translational-rotational temperatures $T$ is investigated computationally 
for conditions of microwave induced plasmas.
Semi-analytic treatment of vibrational relaxation in CO$_2$ in shock tube and acoustic experiments is summarized. 
A state-to-state vibrational kinetics model applied for the simulations 
is benchmarked and adjusted to the relaxation times obtained in gas dynamic experiments.
The governing parameter $Q/n^2_0$ has been introduced, 
where $Q$ is the specific volumetric power coupled in plasma and $n_0$ is the initial number density of CO$_2$. 
The modelling results indicate a rapid increase of the rate of the primary dissociation process 
CO$_2$~+~M~$\to$~CO~+~O~+~M when $Q/n^2_0$ exceeds some critical value. 
A simple analytic calculation of $\left( Q/n^2_0 \right)_{crit}$ is proposed which agrees well with the numerical results. 
At $T$=300~K the estimated $\left( Q/n^2_0 \right)_{crit} \eqsim 6\cdot10^{-40}$~W$\cdot$m$^3$.
\end{abstract}

\noindent{\it CO$_2$, vibrational relaxation, dissociation, microwave plasma, plasma conversion \/}

\maketitle 


\section{Introduction}

Plasma chemical conversion of CO$_2$ into CO is intensively studied 
now as a potential upstream process for production of synthetic fuels and chemicals (Power-2-X)
~\cite{Snoeckx2017,BogaertsNeyts2018,vanRooij2018}. 
Initially those studies were inspired by, presumably, very high energy efficiency 
of the CO$_2$ conversion in microwave plasmas reported in the past~\cite{Legasov1978,Rusanov1981}.
The chemical energy efficiency $\eta_{chem}$ up to 80~\% achieved at relatively low translational-rotational temperatures 
$T\lesssim$1000~K  was ascribed to dissociation from high vibrational states 
in conditions of strong vibrational non-equilibrium~\cite{Capezzuto1976,Legasov1978,Rusanov1981}. 

The reported high efficiencies at low $T$ were not reproduced later. 
In the modern day experiments the highest $\eta_{chem}$ of 40..50~\% 
are observed at $T>$2000~K and are largely explained by thermal quenching \cite{Goede2014,Bongers2017,denHarder2017,DIsa2020}.  
Nevertheless, activating the non-equilibrium conversion mechanism  remains an 
attractive option for increasing the efficiency of plasma processes and 
making them competitive with electrolysis in this respect. 
In the present paper a theoretical and numerical evaluation is given of the conditions 
at which the non-equilibrium process could be triggered. 

The most complete computational description of the CO$_2$ vibrational kinetics and chemistry
is provided by the dedicated state-to-state models~\cite{Kozak2014,Pietanza2020,Armenise2018,Kunova2020,KotovPSST2021}.
Despite their complexity the models developed so far still 
involve a number of uncertainties and do not allow accurate quantitative predictions. 
At the same time, as it will be shown here, already on this stage the 0D vibrational kinetics 
models can be useful in determining the threshold above which the target process could be activated in principle. 
This threshold is defined in terms of the parameter $Q/n^2_0$, where $Q$ is the specific 
(volumetric) power input into  electrons, and $n_0$ is the initial number density of molecules. 

The computational study is performed here with the 2-modes model~\cite{KotovPSST2021} 
where the treatment of the CO$_2$ 
kinetics is simplified by introducing one effective mode for both kinds of symmetric vibrations.
In a series of numerical experiments the rate of the primary dissociation process 
CO$_2$~+~M~$\to$~CO~+~O~+~M is shown to rapidly increase starting from zero 
when $Q/n^2_0$ is increased above some critical value. 
This critical value is found to be not sensitive with respect to model uncertainties. 
Furthermore, the process threshold can be also calculated approximately 
on the basis of a simple semi-empiric balance equation for vibrational energy 
which originates from vibrational relaxation studies in shock tube and sound absorption experiments.
The threshold determined by the semi-empiric calculation agrees well the results of numerical simulations.
The obtained critical value $\left(Q/n^2_0\right)_{crit}$ can be used for analysis and 
planning of experiments aiming at achieving the non-equilibrium vibrational mechanism of CO$_2$ conversion.

Since the rate of vibrational relaxation in CO$_2$ plays a decisive role 
in determining the triggering conditions of the non-equilibrium process 
the first part of the paper focuses specifically on that topic. 
Section~\ref{relaxation:equation:S} summarizes the analytic and semi-analytic models which 
describe the relaxation of vibrational energy in CO$_2$. 
In section~\ref{shock:wave:S} the 2-modes model~\cite{KotovPSST2021} is benchmarked 
against the experimental relaxation times of the gas dynamic experiments.
It is demonstrated that the model can be adjusted such that 
the experimental relaxation times are matched.
The rest of the paper deals with the main subject of the present work. 
Section~\ref{model:runs:s} describes 
the numerical experiments with the calibrated model~\cite{KotovPSST2021} applied to conditions of microwave induced plasmas. 
An approximate approach for calculating the critical values of $Q/n^2_0$ is introduced
in section~\ref{approximate:threshold:S}, and verified with help of the simulation results. 
Last section gives a summary of the main findings and a brief outlook on the outstanding issues.

\section{Analytic treatment of vibrational relaxation in CO$_2$}

\label{relaxation:equation:S}

The basic information about vibrational relaxation
in molecular gases is obtained in shock tube and acoustic experiments~\cite{HerzfeldLitovitz1959,Taylor1969,Simpson1969}. 
Here for compactness the consideration is structured around relaxation in shock waves.
Same relaxation processes take place in sound waves and the relaxation rates deduced from both types of experiments agree~\cite{Taylor1969,Simpson1969}.  
Behind shock wave fronts a very rapid increase of translational temperature is followed by 
its fast equilibration with rotational temperature.  
The characteristic time of translational-rotational relaxation 
is $\sim$10$^{-9}$~sec at atmospheric pressure 
(see~\cite{HerzfeldLitovitz1959}, Section 48, 49). 
The thermal excitation of vibrational states is much slower: 
the typical characteristic times at 1~bar are 10$^{-6}$..10$^{-5}$~sec. At the same time, 
unless the temperatures are not very high, the change of the gas composition 
due to chemical reactions takes place on a significantly longer time scale. 

It is well known that for a gas of diatomic molecules with unchanging number density 
the relaxation of vibrational energy after instant 
increase of translational-rotational temperature $T$ can be described by  
a simple differential equation, see e.g.~\cite{HerzfeldLitovitz1959}, Chapter 19: 
\begin{equation}
 \frac{dE_{vibr}}{dt} =  \frac{1}{\tau}  \left[ E^{eq}_{vibr}\left(T\right) - E_{vibr} \right]
 \label{evibr:relaxation:eq}
\end{equation}
Here $E_{vibr}$ is the average specific vibrational energy per molecule,  
$E^{eq}_{vibr}\left(T\right)$ is the average vibrational energy per molecule at Boltzmann equilibrium with temperature $T$, 
$\tau$ is the relaxation time which depends only on $T$ and on the number density of molecules $n$:
\begin{equation}
\tau^{-1} = R_{10}\left(T\right) n \left( 1 - e^{-\frac{\hbar\omega}{T}} \right) 
\label{diatomic:tau:eq}
\end{equation}
$R_{10}$ is the rate coefficient of transition from the first vibrationally excited state
to vibrational ground state, $\hbar\omega$ is the energy of one vibrational quantum. 
Here and below, unless specified explicitly, the temperatures are always expressed in energy units.  

Equations~(\ref{evibr:relaxation:eq}),~(\ref{diatomic:tau:eq}) are derived on the following assumptions: 
i) the molecules are linear (harmonic) oscillators; ii) the probabilities of vibrational-translational (VT) transitions obey the SSH
(Schwartz-Slawski-Herzfeld) or Landau-Teller relations; iii) the number of vibrational states is infinite.
For CO$_2$ which has 3 modes of oscillations~(\ref{evibr:relaxation:eq}),~(\ref{diatomic:tau:eq}) have to be modified. 
Two more assumptions are added:
iv) exchange of vibrational energy with translational-rotational 
modes proceeds only via transitions between the bending mode oscillations:
\begin{equation}
CO_2\left(v_1 v^l_2 v_3\right) + M \to CO_2\left(v_1, v^{l\pm1}_2\pm1, v_3\right) + M
\label{bending:modes:transition:eq}
\end{equation}
v) exchange of energy between bending modes and stretching modes is very fast. 
Rigorous derivation under assumptions above can be found in~\ref{vibrational:relaxation:CO2:S}.
The resulting equation is same as~(\ref{evibr:relaxation:eq}), 
but with $E_{vibr}$ on the right hand side replaced by $E_{vibr-2}$ - the specific 
vibrational energy per molecule accumulated in bending mode $v^l_2$ only:
\begin{equation}
 \frac{d E_{vibr} }{dt} = R_{10}\left(T\right)n \left( 1 - e^{-\frac{\hbar\omega_2}{T}} \right) 
  \left[E^{eq}_{vibr-2}\left(T\right) - E_{vibr-2} \right]
 \label{evibr:CO2:relaxation:eq}
\end{equation}
\begin{equation}
 E^{eq}_{vibr-2}\left(T\right)  = \frac{2\hbar\omega_2}{  e^{\frac{\hbar\omega_2}{T}} - 1 }
 \label{evibr2:harmonic:eq}
\end{equation}
here $\hbar\omega_2$ is the fundamental energy of bending oscillations $v_2$.
$E^{eq}_{vibr-2}\left(T\right)$ is calculated by applying the known formula for harmonic oscillators,
see e.g.~\cite{LandauLifshitz5}, equation (49.3) there. Factor 2 takes into account that the mode $v_2$ of CO$_2$ 
is double degenerate. $R_{01}$ is the rate coefficient of the process: 
\begin{equation}
CO_2\left(0 1^1 0\right) + M \to CO_2\left(00^00\right) + M
\label{bending:modes:transition10:eq}
\end{equation}

In~\cite{MontrollShuler1957} it had been shown that when a system of linear oscillators 
initially has Boltzmann distribution with temperature equal to the translational-rotational temperature $T$, 
and $T$ is instantly increased, then vibrational relaxation towards final equilibrium 
state proceeds via a sequence of Bolzmann distributions with vibrational temperature $T_{vibr} \le T$. 
The result of~\cite{MontrollShuler1957} is not directly applicable to 
CO$_2$, but numerical experiments with state-to-state model~\cite{KotovPSST2021} presented below in section~\ref{shock:wave:S} 
confirm its validity in that case as well. 
For the Boltzmann vibrational distribution equation~(\ref{evibr:CO2:relaxation:eq}) is transformed into equation for $T_{vibr}$ which can be easily integrated:
\begin{equation}
  \frac{d T_{vibr} }{dt}= 
  \frac{ R_{10}\left(T\right) n \left( 1 - e^{-\frac{\hbar\omega_2}{T}} \right) }{ c_{vibr}\left(T_{vibr} \right) }
  \left[E^{eq}_{vibr-2}\left(T\right) - E^{eq}_{vibr-2}\left(T_{vibr}\right)  \right] 
   \label{Tvibr:CO2:relaxation:eq}
\end{equation}
Here $c_{vibr}\left(T_{vibr} \right) =  \frac{d E_{vibr} }{ d T_{vibr}}$ 
is the (dimensionless) heat capacity due to vibrational modes which is calculated by using the formulas
for linear oscillators, see e.g.~\cite{LandauLifshitz5}, equation (49.4):
\begin{equation}
c_{vibr}\left(T_{vibr} \right) = c_1 + 2 c_2 + c_3, \quad 
c_i = \frac{ \left( \hbar\omega_i \right)^2 e^{\frac{\hbar\omega_i}{T_{vibr}}} }{ T^2_{vibr} \left( e^{\frac{\hbar\omega_i}{T_{vibr}}} - 1 \right)^2 } 
\end{equation}
$\hbar\omega_i$ are the fundamental energies of the 3 modes of CO$_2$ oscillations. 

The assumption of Bolzmann distribution of vibrational levels allows to derive one more equation where the rate of vibrational relaxation appears explicitly
- by transforming~(\ref{evibr:CO2:relaxation:eq}) into equation for $E_{vibr-2}$:
$$
 \frac{d E_{vibr} }{dt} = \frac{d E_{vibr} }{d E_{vibr-2} } \frac{d E_{vibr-2} }{dt}  = \frac{ \frac{d E_{vibr}}{dT_{vibr}} }{  \frac{d E_{vibr-2}}{dT_{vibr}} }\frac{d E_{vibr-2} }{dt} =
  \frac{ c_{vibr}\left( T_{vibr} \right) }{ 2c_2\left( T_{vibr} \right) }\frac{d E_{vibr-2} }{dt}
$$
\begin{equation}
\frac{d E_{vibr-2} }{dt} = \frac{ 2 c_2 \left( T_{vibr} \right) }{  c_{vibr} \left( T_{vibr} \right)   }  R_{10}\left(T\right)n \left( 1 - e^{-\frac{\hbar\omega_2}{T}} \right) 
  \left[E^{eq}_{vibr-2}\left(T\right) - E_{vibr-2} \right]
 \label{evibr2:CO2:relaxation:eq}
\end{equation}
One can see that the rate of relaxation (subsequently, the instant relaxation time) is not independent of $T_{vibr}$, but this dependency is weak. 
The factor $\frac{ 2 c_2}{ c_{vibr} }$ only changes from 1 at low $T_{vibr}$ where $c_2\gg c_1,\,c_3$ 
(because $\hbar\omega_2 < \hbar\omega_1,\, \hbar\omega_3$) to $\frac{ 2 c_2}{ c_{vibr} } = \frac12$ 
at high $T_{vibr}$ where $c_1,\, c_2,\, c_3 \to 1$.

Equation~(\ref{evibr2:CO2:relaxation:eq}) suggests the following approximation. 
One can assume that the total vibrational energy $E_{vibr}$
decays with exactly same rate as $E_{vibr-2}$, and then apply 
equation~(\ref{evibr:relaxation:eq}) with $\tau$ defined as:
\begin{equation}
 \tau^{-1} =   \frac{ 2 c_2 \left( \bar T \right) }{  c_{vibr} \left( \bar T \right)   }  R_{10}\left(T\right)n \left( 1 - e^{-\frac{\hbar\omega_2}{T}} \right) 
 \label{approximate:tau:eq}
\end{equation}
Here $\bar T$ is some average temperature, e.g. $\bar T = \frac12 \left(T_0 + T \right)$ where $T_0$ is the 
initial Boltzmann temperature before the vibrational relaxation starts. Of note is that a relation similar 
to~(\ref{approximate:tau:eq}) appears also in~\cite{Taylor1969}, equation (24) there.

To estimate the accuracy of this approximation a comparison is made 
with numerical solution of equation~(\ref{Tvibr:CO2:relaxation:eq}) with constant $T$.
Both solutions are expressed in terms of the dimensionless time:
\begin{equation}
t' = t\cdot  R_{10}\left(T\right) n \left( 1 - e^{-\frac{\hbar\omega_2}{T}} \right)
\label{dimensionless:time:eq}
\end{equation}
That is, the explicit dependence on $n$ and $R_{10}$ is eliminated. 
The temperature $T_{vibr}$ obtained by solving~(\ref{Tvibr:CO2:relaxation:eq}) is 
translated into $E_{vibr}$ by the linear oscillator formula:
\begin{equation}
E_{vibr} \left( T_{vibr} \right) = \frac{\hbar\omega_1}{  e^{\frac{\hbar\omega_1}{T_{vibr}}} - 1 } +  \frac{2\hbar\omega_2}{  e^{\frac{\hbar\omega_2}{T_{vibr}}} - 1 } 
         + \frac{\hbar\omega_3}{  e^{\frac{\hbar\omega_3}{T_{vibr}}} - 1 }
\label{evibr:CO2:harmonic:eq}         
\end{equation}
The fundamental energies of the CO$_2$ oscillations~\cite{Teffo1992}: 
$\hbar\omega_1$=0.16783~eV, $\hbar\omega_2$=0.083427~eV, $\hbar\omega_3$=0.29710~eV. 
The time evolution $E_{vibr}\left(t'\right)$ is compared with the exponential function:
\begin{equation}
E^\tau_{vibr}\left(t\right) = E_{vibr}\left(T\right) + \left[  E_{vibr}\left(T_0\right)  -  E_{vibr}\left(T\right) \right] e^{-t/\tau} 
\label{evibr:exp:eq}
\end{equation}
where $t$ is replaced by $t'$ and $\tau$ is replaced by $\tau' = \frac{  c_{vibr} \left( \bar T \right) }{ 2 c_2 \left( \bar T \right) }$, 
see~(\ref{approximate:tau:eq}) and~(\ref{dimensionless:time:eq}). 
The relative difference $\epsilon$ between two solutions is defined as:
\begin{equation}
 \epsilon = \max_{t'}{ \left[  \frac{\left| E_{vibr}\left(t'\right) - E^\tau_{vibr}\left(t'\right) \right|}{  E_{vibr}\left(t'\right)  } \right] }
 \label{epsilon:eq}
\end{equation}
where the maximum is taken over the time interval from $t'=0$ to $t'=10\cdot \tau'$. 

For the test with $T_0$=300~K and  $\bar T = \frac12 \left(T_0 + T \right)$ the quantity $\epsilon$ 
is found to be always $\le$5~\% for $T\le$2500~K. 
This outcome means that, provided that all the assumptions 
underlying~(\ref{Tvibr:CO2:relaxation:eq}) are fulfilled,   
the time evolution of $E_{vibr}$ can be approximated with good accuracy by equation~(\ref{evibr:relaxation:eq})
with $\tau$ defined by~(\ref{approximate:tau:eq}). This characteristic time, in turn, 
depends on the rate coefficient of only one process~(\ref{bending:modes:transition10:eq}). 
The coefficient $R_{10}$ then serves as a single parameter which fits the 
vibrational kinetics model into experimental data. It will be shown in the next section that this 
result holds as well when equation~(\ref{Tvibr:CO2:relaxation:eq}) is replaced by state-to-state simulations. 

\section{Benchmark of the CO$_2$ vibrational kinetics for conditions of gas-dynamic experiments}

\label{shock:wave:S}

The equations introduced the previous section can be used to calibrate a state-to-state model 
of the CO$_2$ vibrational kinetics 
such that it reproduces 
the experimental relaxation times in a situation which mimics conditions behind a shock front. 
This will be demonstrated here with the 
so called '2-modes' model described in~\cite{KotovPSST2021}. 
This is a coarse-grained model where elementary vibrational states $CO_2\left(v_1 v^l_2 v_3\right)$ are gathered 
into 'combined' symmetric-asymmetric states 
$CO_2\left[v_s, v_a\right]$;
$v_a=v_3$, and $v_s=2v_1+v_2$ is a good quantum number which combines symmetric stretching 
and bending modes. The model includes dissociation of CO$_2$ from the 'unstable' states 
with vibrational energies exceeding the dissociation limit of 5.5~eV. 
The rates of vibrational transitions between combined states are calculated 
by applying scalings based mainly on the Herzfeld (SSH) theory~\cite{Herzfeld1967}. 
The absolute values of the rate coefficients of transitions between symmetric and asymmetric modes are adjusted to the 
laser florescence data. The same applies to vibrational-vibrational (VV) transitions between 
asymmetric modes, the rates of VV-transitions between symmetric modes are purely theoretical. 
Adjustment (calibration) of the rates of vibrational-translational (VT)
transfer~(\ref{bending:modes:transition:eq}) is discussed below.

\begin{figure}
\includegraphics[width=7cm]{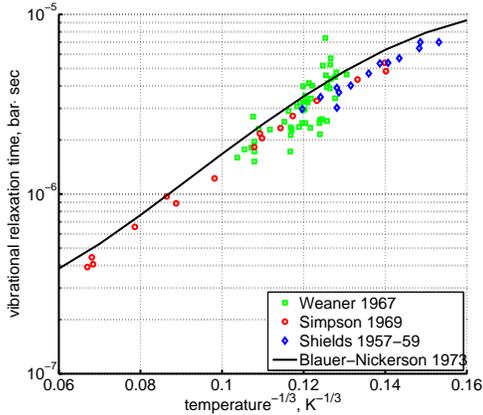} 
\caption{Experimental characteristic time of vibrational relaxation $\tau$, 
see equation~(\ref{evibr:relaxation:eq}), at pressure 1~bar. 
'Weaner 1967' is~\cite{Weaner1967}, 'Simpson 1969'is~\cite{Simpson1969} 'Shields 1967-59’ is~\cite{Shields1957,Shields1959} as 
reproduced in~\cite{Simpson1969}. 
Solid line is the fit from~\cite{BlauerNickerson1973} }
\label{tauVT:Fig} 
\end{figure}

The experimental data on VT-transfer in CO$_2$ had been commonly published 
in form of relaxation times 
$\tau$ defined in terms of equation~(\ref{evibr:relaxation:eq}). 
The experimental $\tau$s could be reduced to functions of 
translational-rotational temperature $T$ only~\cite{Taylor1969,Simpson1969}.
The most recent review found on the subject is the publication~\cite{JolyRobin1999}. 
By comparing that paper with the older report~\cite{BlauerNickerson1973} 
one finds no references on process~(\ref{bending:modes:transition:eq}) 
with M=CO$_2$ which had not been taken into account in~\cite{BlauerNickerson1973}.
Therefore, here the rate coefficient of process~(\ref{bending:modes:transition10:eq}) 
as a function of $T$ is taken from~\cite{BlauerNickerson1973}
(process $\left(0 1^1 0\right) \to \left(0 0^0 0\right)$, M=CO$_2$, in Table IIIa there)
as the best fit of the all available experimental data. 
Apparently, this rate coefficient was obtained by applying equation~(\ref{diatomic:tau:eq}) 
with $\hbar\omega = \hbar\omega_2$ rather than equation~(\ref{approximate:tau:eq}). 
The correctness of this assumption is confirmed by translating the rate 
coefficient back into $\tau\left(T\right)$, and comparing the result 
with selected primary publications on the relaxation time measurements. 
This comparison is shown in figure~\ref{tauVT:Fig} where  $\tau\left(T\right)$
'reverse engineered' as described above ('Blauer-Nickerson 1973') is 
plotted together with results of the shock tube~\cite{Weaner1967,Simpson1969} and 
sound absorption~\cite{Shields1957,Shields1959} experiments. The agreement is very good. 

The rate coefficient of process~(\ref{bending:modes:transition10:eq}) applied here 
is calculated by using equation~(\ref{approximate:tau:eq})  
with $\tau$ derived from the rate coefficient $R^{BN73}_{10}$ of that process  in~\cite{BlauerNickerson1973}. That is:
\begin{equation}
R_{10}\left(T\right) = \frac{c_{vibr}\left(\bar T\right) }{2 c_2\left(\bar T\right) } R^{BN73}_{10}\left(T\right),\quad
\bar T = \frac12\left(300\;{\rm K} + T  \right)
\label{R10:vs:BN73:eq}
\end{equation}
The coefficient $R^{BN73}_{10}$ corrected in that way can be fitted by the equation:
\begin{equation}
R_{10}\left(T\right) = 4.48\cdot10^{-13}\exp{\left( -198.8 T^{-1/3} + 517.6 T^{-2/3} \right)}
\label{R10:fit:eq}
\end{equation}
Here $T$ is in Kelvin and the resulting $R_{10}$ is in m$^3$/s. For $T$ from 250~K to 
2000~K~(\ref{R10:fit:eq}) approximates~(\ref{R10:vs:BN73:eq}) with maximum relative deviation 0.5~\%. 
The magnitude of the correction factor was already discussed in section~\ref{relaxation:equation:S}: 
the original and corrected rate coefficients are nearly equal at low $T$,  
and $R_{10}$ is twice as large as 
$R^{BN73}_{10}$ at high $T$ ($\ge$2000~K). The rate coefficients of the transitions from 
higher excited states are obtained from $R_{10}$ by applying SSH scaling as described in~\cite{KotovPSST2021}. 

The 2-modes model is benchmarked by applying the same test problem as in the previous section.
Initial gas has Boltzmann vibrational distribution with $T_0$=300~K, 
the temperature $T>T_0$ is prescribed and fixed. 
The system relaxes by thermal excitation of vibrational states 
until the equilibrium vibrational distribution is reached which 
corresponds to the temperature $T$. For $T\le$2500~K considered here the change of the CO$_2$ number density due to 
dissociation on the time-scale of the test is in all cases $<$0.01~\%.

In a very general sense a state-to-state model can be written as initial 
value problem for the set of ordinary differential equations: 
\begin{equation}
\frac{d n_k}{d t} = \sum_{i,j} a^k_{ij} n_i n_j 
\label{state2state:eq}
\end{equation}
Where $n_k$, $n_i$, $n_j$ are the number densities of individual species - excited states, 
and coefficients $a^k_{ij}$ are functions of the temperature $T$ only
. 
It can be easily shown that changing of the gas pressure $p$ at fixed $T$ does not change 
the shape of the solution of~(\ref{state2state:eq}). Indeed, replacing densities with concentrations $c$: 
$n_k=c_kn_0$, where $n_0$ is the total initial density of molecules, yields
\footnote{technically, in the model~\cite{KotovPSST2021} 
always the initial density $n_0$ is used as the density of species M, 
therefore, equation~(\ref{state2state:eq}) should also contain terms with $n_in_0$; 
this peculiarity is ignored because it does not change the final result}:
$$
\frac{1}{n_0} \frac{d c_k}{d t} = \sum_{i,j} a^k_{ij} c_i c_j 
$$
The solution of this equation written in terms of $n_0 t$ - the integration time scaled proportional 
to $n_0$ - depends only on $T$. 
Therefore, all the reference model runs here are made for one nominal pressure $p$=1~bar. 
Two test with $p$=0.1~bar and $p$=10~bar yield, as expected, exactly same results within 
relative difference 10$^{-7}$ when the time-traces 
$E_{vibr}\left(n_0t\right)$ are compared.

\begin{figure}
\includegraphics[width=7cm]{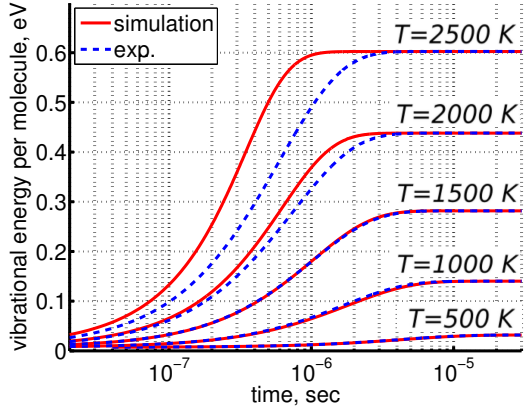}  
\caption{Comparing the time evolution of vibrational energy $E_{vibr}$   
in the test problem of section~\ref{shock:wave:S} simulated by the 2-modes model~\cite{KotovPSST2021}
(solid lines) and approximated by the exponential function~(\ref{evibr:exp:eq}) 
with $\tau$ defined by~(\ref{approximate:tau:eq}) (dashed lines). 
$T$ is the fixed translational-rotational temperature}
\label{benchmark:Fig} 
\end{figure}

The time evolution of specific vibrational energy $E_{vibr}$ calculated by the 2-modes model 
is compared with exponential function~(\ref{evibr:exp:eq}) 
with $\tau$ defined by~(\ref{approximate:tau:eq}). To remind, since $R_{10}$ is 
defined by~(\ref{R10:vs:BN73:eq}) this $\tau$ is the fit of the primary experimental data.  
Selected time-traces of $E_{vibr}$ are shown in figure~\ref{benchmark:Fig}. 
One can see a very good agreement up to $T$=1500~K, the agreement deteriorates at higher temperatures. 
Quantitative comparison of the numerical solution and the analytic formula is given in table~\ref{benchmark:Table}
in terms of relative deviation $\epsilon$ defined by~(\ref{epsilon:eq}). 
The meaning of the parameter $p^{max}_{SSH}$ will be explained below. 
One can see that for $T\le$1500~K the exponential function fits the numerical solution within 4~\%. 
That is, the conclusion on the applicability of the simple formula~(\ref{approximate:tau:eq}) 
made in the previous section is confirmed by a state-to-state model. 
Or, the other way around, the test demonstrates that in conditions which 
mimic those behind a shock front the state-to-state model 
calibrated as prescribed by equation~(\ref{approximate:tau:eq}) 
yields at $T\le$1500~K (to some extent up to 2000~K)
the same relaxation behavior as the empiric equation~(\ref{evibr:relaxation:eq}).

\begin{table}
\caption{Comparison of $E_{vibr}\left(t\right)$ in the test problem of
section~\ref{shock:wave:S}  simulated by the 2-modes model~\cite{KotovPSST2021} 
and approximated by~(\ref{evibr:exp:eq}),~(\ref{approximate:tau:eq}) 
in terms of the relative error $\epsilon$ defined by~(\ref{epsilon:eq}) }
\begin{tabular}{c|cc|cc}
\hline
       & \multicolumn{2}{c|}{ $p^{max}_{SSH}$=1 }  & \multicolumn{2}{c}{ $p^{max}_{SSH}$=0.1 } \\
\hline 
$T$, K & $\epsilon$, \% & $\eta_{V1}$ \% & $\epsilon$, \%  & $\eta_{V1}$ \% \\ 
\hline
350  &  0.2 (0.2) & 100.0                      & 0.2  & 100.0 \\ 
400  &  0.6 (0.5) & 99.9                       & 0.6  & 99.9  \\ 
500  &  1.6 (1.5) & 99.8                       & 1.6  & 99.8 \\ 
600  &  2.6 (2.6) & 99.6                       & 2.6  & 99.6 \\ 
800  &  3.9 (3.9) & 98.9                       & 3.9  & 98.9 \\ 
1000 &  4.0 (4.0) & 97.7                       & 4.0  & 97.7 \\ 
1200 &  3.1 (3.2) & 95.5                       & 3.1  & 95.5 \\ 
1500 &  1.0 (0.6) & 89.6                       & 1.0  & 89.6 \\ 
2000 &  12.2 (10.4) & 71.9                     & 12.2 & 72.0 \\ 
2500 &  31.9 (28.4) & 48.7                     & 31.3 & 50.3 \\ 
\hline
\end{tabular}
\label{benchmark:Table} 
\end{table}

The numerical simulations allow to verify the 
assumptions behind the equations~(\ref{Tvibr:CO2:relaxation:eq}) and~(\ref{approximate:tau:eq}).  
In particular, of the linear oscillators assumption which is only applied 
in the 2-modes model for calculation of 
matrix elements of transitions. The vibrational energies are calculated with the full 
anharmonic expression~\cite{Teffo1992}. 
To estimate the importance of anharmonicity the simulations were repeated 
with energies calculated in linear (harmonic) approximation. 
The resulting values of $\epsilon$ are shown in the second column of
table~\ref{benchmark:Table} in parentheses, 
the difference in the outcome of the benchmark is negligible. 

The most non-obvious assumption underlying the derivation of~(\ref{Tvibr:CO2:relaxation:eq}) 
is that of the dominance of transition~(\ref{bending:modes:transition:eq}) 
in the thermal excitation of vibrational energy.
For its verification in 2-modes calculations the
relative contribution of process~(\ref{bending:modes:transition:eq}) in the total energy exchange
is calculated:
$$
\eta_{V1} = \frac{ \int_t Q_{V1}\left(t\right)dt }{  \int_t Q_{tot}\left(t\right)dt } 
$$
here $Q_{tot}\left( t \right)$ is the total power transferred from translational-rotational to vibrational modes, 
$Q_{V1}\left(t\right)$ is the power transferred by the process~(\ref{bending:modes:transition:eq}) only, 
the integral is calculated over the whole time interval where the equations are solved.
One can see from table~\ref{benchmark:Table} that $\eta_{V1}\ge$90~\% for $T\le$1500~K, and this is exactly the 
temperature range where the agreement between the 2-modes simulations and 
the semi-empiric fit is good. Above 2000~K the assumption 
of the dominance of~(\ref{bending:modes:transition:eq}) does not hold any more, 
and the relaxation equations of section~\ref{relaxation:equation:S} are not applicable. 
At high $T$ the transfer of translational-rotational energy into vibrational modes
is taken over by inter-mode transitions $CO_2\left(v_1,v^l_2,v_3\right) + M \to CO_2\left(u_1,u^m_2,v_3\pm1\right) + M$
(processes V2 in~\cite{KotovPSST2021}). At $T\ge$3000~K this is the dominant mechanism of thermal 
excitation of vibrational energy. This conclusion drown for high temperatures has to be taken with caution
because besides the shortcomings of the SSH scaling which will be discussed below 
the absolute  values of the rate coefficients of the inter-mode transitions applied in the present model 
are based on the laser fluorescence data available only for $T\le$1000~K.

The SSH theory~\cite{Herzfeld1967}
applied in 2-modes model~\cite{KotovPSST2021} to scale the probabilities of vibrational transitions 
from low to high excited states is a first order perturbation theory.
It is known that this theory can grossly overestimate the 
transition probabilities for very high excited states and at high temperatures. 
To cope with that issue in the model 
the transition probabilities treated by the SSH theory are not allowed to be larger than the prescribed  parameter $p^{max}_{SSH}$.
The nominal value of this parameter is 1. In order to check the influence of $p^{max}_{SSH}$
the test of the present section was repeated with $p^{max}_{SSH}$=0.1.
The results are shown in the two last columns of table~\ref{benchmark:Table}. 
The difference is only visible starting from $T$=2000~K, and does not affect the conclusions drawn above. 

\begin{figure}
\includegraphics[width=7cm]{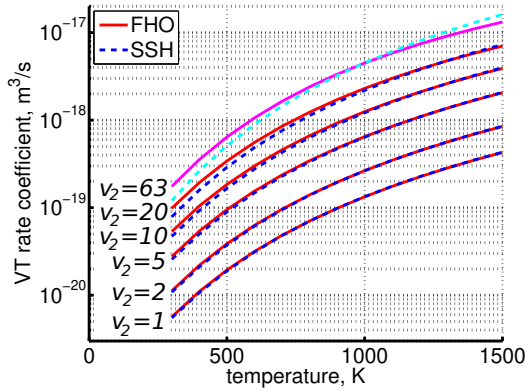}
\caption{Rate coefficients of the process~(\ref{bending:modes:transition:FHO:eq})
taken from the set of FHO calculations ~\cite{VargasJPC2021,daSilva2018,STELLAR} (solid lines) 
compared with the coefficients calculated according to the SSH theory~\cite{Herzfeld1967} }
\label{FHO:Fig} 
\end{figure}

The shortcoming of the SSH theory is solved in the non-perturbative Forced Harmonic Oscillators (FHO) 
model~\cite{Adamovich1998}. Recently, FHO calculations of the probabilities 
of vibrational transitions in CO$_2$ 
became available~\cite{VargasJPC2021,daSilva2018,STELLAR}.
A comparison between the rate coefficients $R$ 
of VT-transition~(\ref{bending:modes:transition:eq}) 
calculated by the SSH and FHO methods has shown that 
the FHO correction has only a small impact on the calculated rate of that process.
This comparison was performed for the set of the FHO rate coefficients of the transitions: 
\begin{equation}
 CO_2\left(0 v_2^{v_2} 0 \right) + M \to CO_2\left(0 \left(v_2-1\right)^{v_2-1} 0 \right) + M,\quad M = CO_2
\label{bending:modes:transition:FHO:eq}
\end{equation}
taken from~\cite{STELLAR}, file \verb'K_CO2v2_CO2_VT.mat', following the description in~\cite{daSilva2018}.
The SSH calculations are described in~\cite{KotovPSST2021,KotovJPB2020}, 
the vibrational energies of the molecules are calculated according to~\cite{Teffo1992}.
The rate coefficients of selected transitions calculated by the two models are plotted in figure~\ref{FHO:Fig}. 
Both SSH and FHO calculations are normalized such that for $v_2$=1 
the rate coefficient equals to that of the  
process $\left(0 1^1 0\right) \to \left(0 0^0 0\right)$ in Table IIIa of~\cite{BlauerNickerson1973}.
One can see that in the temperature range from 300 to 1500~K the deviation between SSH and
FHO scalings is small. For $v_2\le$10 which corresponds to total vibrational 
energy of the molecule $\le$0.84~eV the relative difference 
$\epsilon_{SSH}=\left|R_{FHO}-R_{SSH}\right|/R_{FHO}$ is always $\le$12~\%. 
The upper boundary of $\epsilon_{SSH}$ is increased for higher $v_2$, e.g. 
for $v_2$=20 (energy 1.70~eV) $\epsilon_{SSH}\le$20~\%, but even for 
the highest level $v_2$=63 (energy 5.54~eV) $\epsilon_{SSH}$ stays smaller than 32~\%.

The FHO model also allows to calculate the probabilities of multi-quantum transitions 
which are not accounted for in first order perturbation theories. 
Comparison between the rate coefficients of the single-quantum transitions~(\ref{bending:modes:transition:FHO:eq})
and the corresponding two-quantum rates 
in the data set~\cite{STELLAR} shows that at room temperature $T$=300~K the 
two-quantum transitions are always by more than a factor 100 less probable. 
The importance of multi-quantum transitions is increased at higher $T$. 
However, the ratio $R_{1-q}/\left(2R_{2-q}\right)$ 
- where $R_{1-q}$ is the rate of a single-quantum transition,
and $R_{2-q}$ is the rate of the two-quantum transition from the same $v_2$ - 
stays $>$10 at $T\le$1000~K even for the largest $v_2$=63 ($>$5 at $T\le$1500~K).
With decreasing $v_2$ the ratio $R_{1-q}/\left(2R_{2-q}\right)$ is increased.

For transitions other than~(\ref{bending:modes:transition:eq}), 
in particular for vibrational-vibrational (VV) transfer, the difference between FHO and SSH scalings can be much large.
As has been shown above, at $T<$2000~K those other transitions are relatively 
unimportant for thermal excitation of vibrational energy. 
At the same time, the inter-mode and VV transitions 
(see~\cite{KotovPSST2021}, Table 1 there) 
are the processes which govern fast energy exchange between different kinds of  
vibrational modes and bring them to Boltzmann distribution with single temperature $T_{vibr}$.
Although the absolute values of their coefficients 
are adjusted to experimental data the SSH scaling can lead 
to very inaccurate values for high excited states. 
Nevertheless, as long as those processes in the model are fast enough 
this inaccuracy may not have large impact on the final result.
Strong coupling between different types of vibrations in the 2-modes simulations  
agrees with the fact that in shock tube experiments only one or two very close relaxation 
times were detected~\cite{Taylor1969,Simpson1969}. 

\section{Simulation of the non-equilibrium vibrational dissociation of CO$_2$ in microwave plasmas}

\label{model:runs:s}

The model~\cite{KotovPSST2021} with vibrational relaxation rate benchmarked as described 
in the previous section is applied to investigate the onset of the non-equilibrium 
vibrational dissociation in CO$_2$. The same approach to modelling of the electron impact 
excitation of vibrational states as in~\cite{KotovPSST2021} is adopted here 
with one important modification. Instead of degree of ionization 
the total specific (volumetric) power input $Q$ into electrons is prescribed and fixed 
in a model run. In practice this means that the electron density $n_e$ 
is calculated as:
\begin{equation}
 n_e = \frac{Q}{\sum_i R_{ei} E_i n_i}
 \label{ne:Q:eq}
\end{equation}
where $n_i$ is the number density of molecular species (excited state) $i$, 
$R_{ei}$ is the total rate coefficient of all the electron impact 
processes with species $i$, 
and $E_i$ is the energy transferred on average from electrons in each collision with $i$. 

The set of master equations solved by the model in the discharge zone $Q>0$ can be written in the 
same form as~(\ref{state2state:eq}) with the electron excitation term added:
$$
\frac{d n_k}{dt} =  \sum_{i,j} a^k_{ij}\left(T\right)n_i n_j + \sum_i R^k_{ei}\left(T_e\right) n_e n_i
$$
Substituting~(\ref{ne:Q:eq}) and introducing concentrations $c_k$: $n_k = c_k n_0$ and the 
Specific Energy Input (SEI) per one initial molecule 
$\epsilon = Qt/n_0$ brings the equations above to the form:
\begin{equation} 
\frac{dc_k}{d\epsilon} = \frac{n^2_0}{Q} \sum_{i,j} a^k_{ij}\left(T\right) c_i c_j + 
\frac{ \sum_i R^k_{ei}\left(T_e\right) c_i}{ \sum_i R_{ei}\left(T_e\right) E_i\left(T_e\right) c_i }
\label{concentration:master:eq}
\end{equation}
The coefficients $R_{ei}$, $R^k_{ei}$, $E_i$ are functions of the electron temperature $T_e$.
As one can see from~(\ref{concentration:master:eq}) 
for the constant $T$ and $T_e$  the solution expressed in terms of $c_i$ and
$\epsilon$ depends only on the parameter $Q/n^2_0$.  

The rate coefficients $R_{ei}$ 
are calculated here assuming the maxwellian electron energy distribution
which is approximately valid for conditions of microwave induced plasmas, see~\cite{Silva2014,Capitelli2017,Pietanza2020}.
A more accurate calculation of the energy distribution function
was considered to be superfluous since no excited state resolved data 
are readily available for the electron impact transitions anyway, thus, only a very basic model 
of those processes is applied. Only one quantum transitions are taken into account and 
same rate coefficients are applied for all excited states, see~\cite{KotovPSST2021}.
The modelling studies with detailed electron kinetics~\cite{Capitelli2017,Pietanza2020} have 
shown that in conditions of microwave plasmas at not too low pressures the kinetic energy acquired by electrons 
predominantly goes into vibrational excitations.
The results of~\cite{Pietanza2020} suggest that this statement is valid at least at 20~Torr.
Therefore, for this kind of gas discharges 
the quantity $Q$ of the present model which does not include the power spent to any other kinds of electron processes 
(such as dissociation or ionization) can be taken as approximately equal 
to the total power coupled in plasma. The electron temperature $T_e$=1~eV used in the reference simulations here 
reflects the typical average electron energy calculated for conditions in question.

\begin{figure}
\subfloat[$T$=300~K]{ 
 \includegraphics[width=7cm]{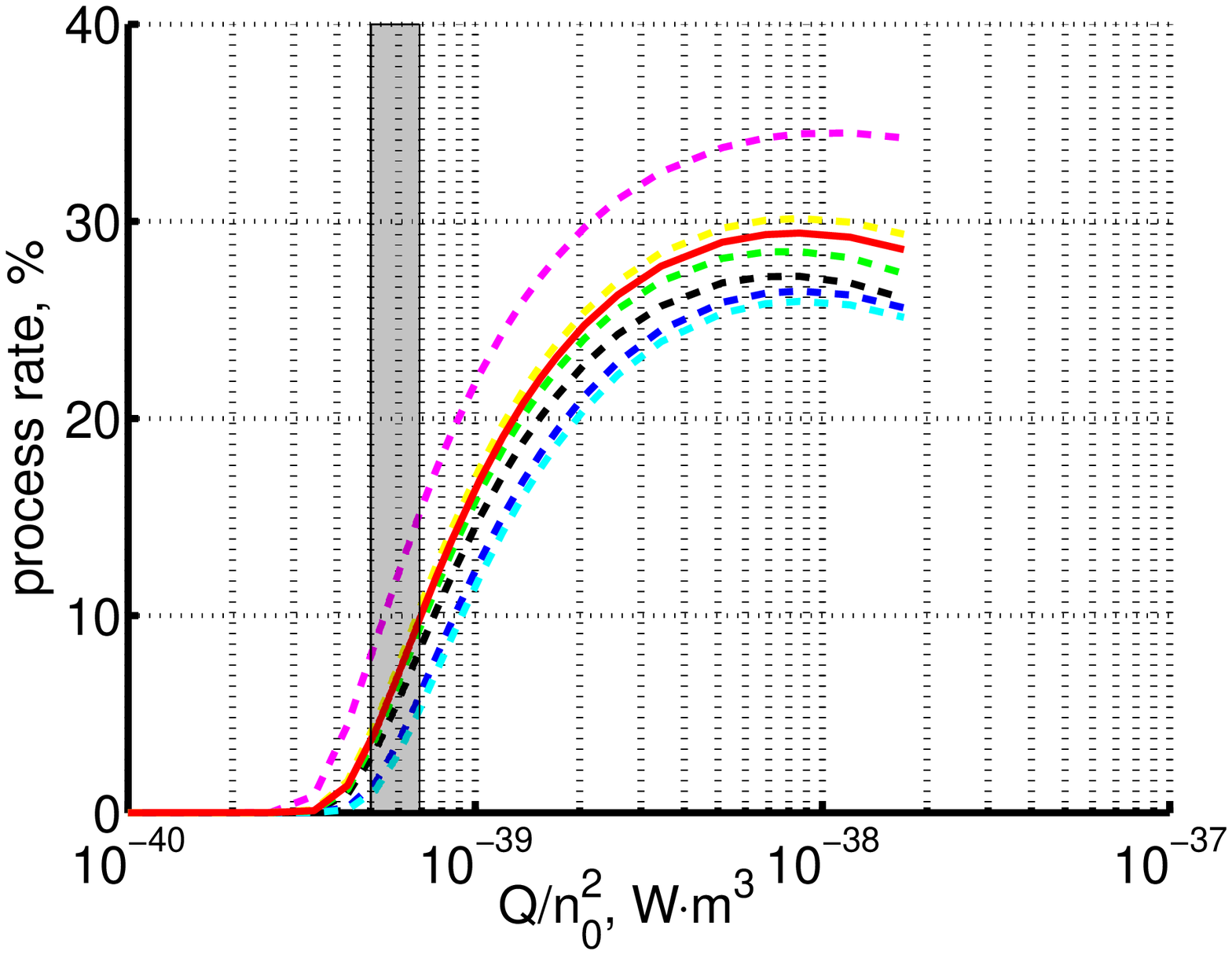} 
 \label{Qn2:a:Fig} }
\subfloat[$T$=500~K]{ 
 \includegraphics[width=7cm]{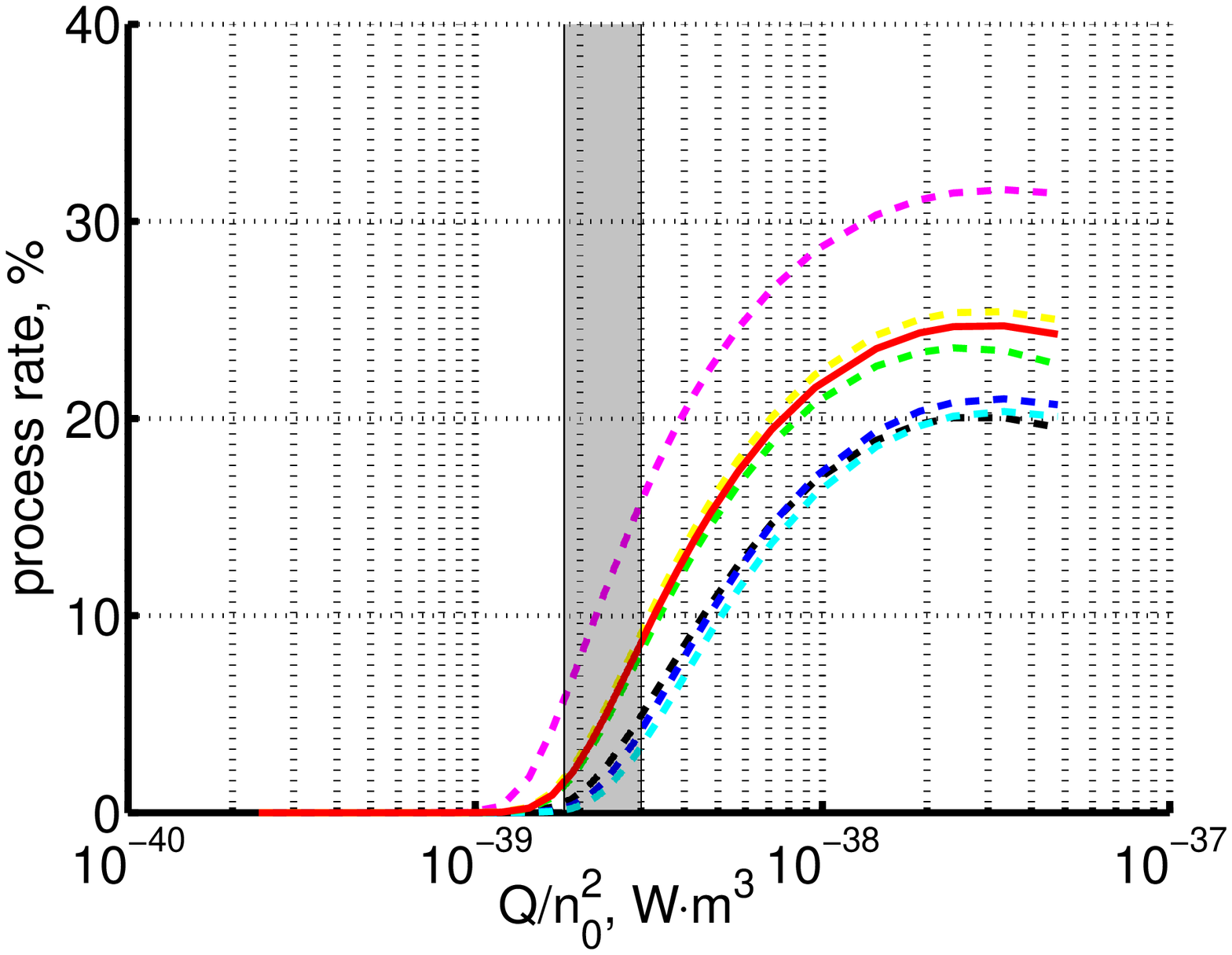} } \\
\subfloat[$T$=800~K]{ 
 \includegraphics[width=7cm]{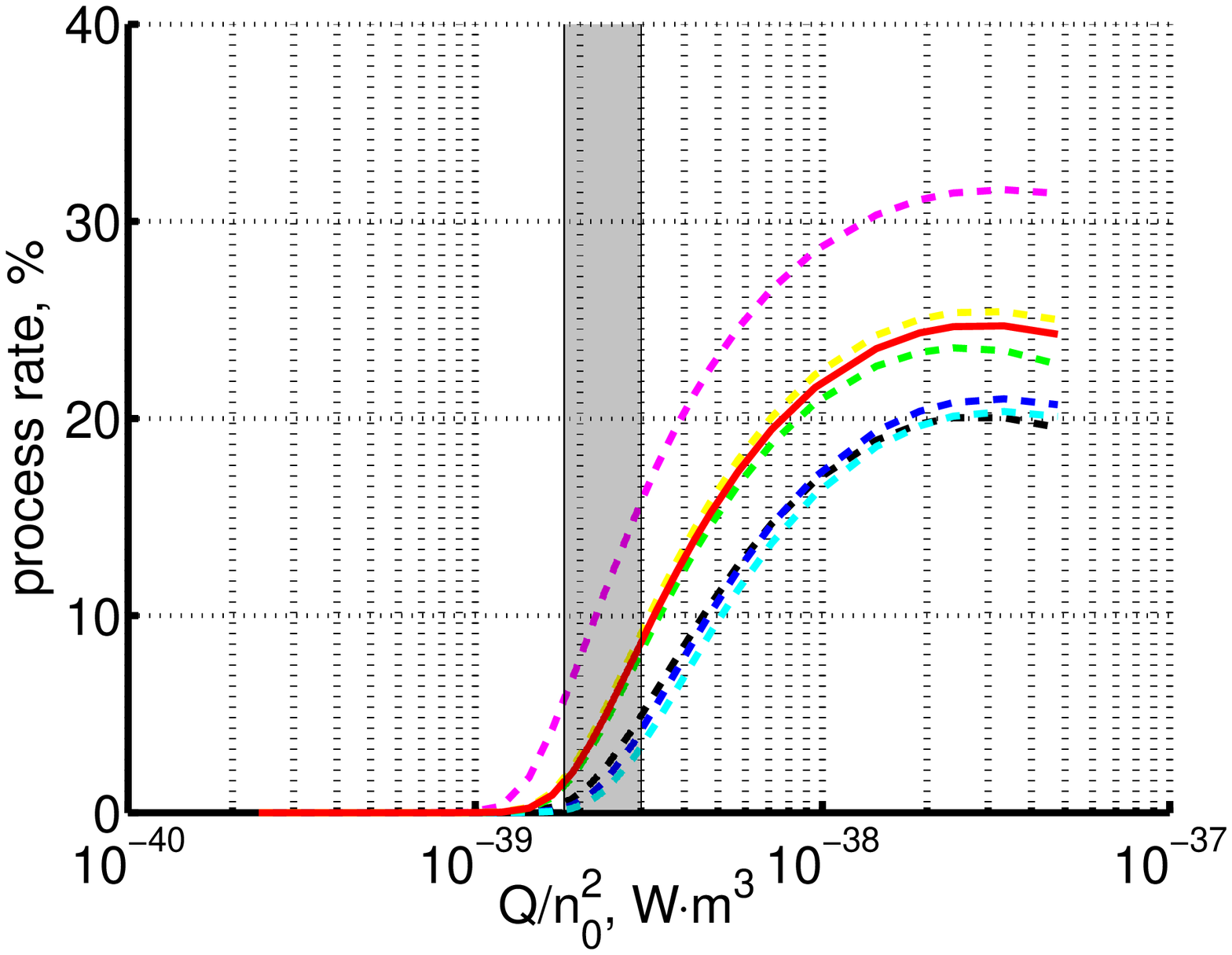} }
\caption{'Process rate' $\chi$ defiend by~(\ref{process:rate:eq}) as a function 
of parameter $Q/n^2_0$, see equation~(\ref{concentration:master:eq}), calculated by the 
2-modes model~\cite{KotovPSST2021} for fixed translational-rotational temperature $T$. 
Solid line is the reference model, dashed lines are the simulations with  
uncertain parameters varied, see section~\ref{model:runs:s}.
Shaded rectangles is the approximate calculation of $\left(Q/n^2_0\right)_{crit}$, section 5}
\label{Qn2:Fig} 
\end{figure}

The outcome of the model calculations for selected fixed temperatures $T$ 
are shown in figure~\ref{Qn2:Fig}. 
The results are presented in terms of $Q/n^2_0$ and the 'process rate' $\chi$ defined as the 
fraction of initial CO$_2$ molecules which dissociate in the 
process CO$_2$~+~M~$\to$~CO~+~O~+~M. 
For all model runs this quantity is taken at the time instants which 
correspond to $\epsilon$=3~eV:
\begin{equation}
\chi = 1 - \frac{n_{CO_2}\left(\epsilon=3~{\rm eV}\right)}{n_{CO_2}\left(\epsilon=0\right)}
\label{process:rate:eq}
\end{equation}
Conversion in the afterglow ($Q$=0) is not taken into account since in the previous work~\cite{KotovPSST2021} 
the conversion on that phase was found to be insignificant. 
The chosen final value of $\epsilon$ is close to the net enthalpy change of the total process  
CO$_2$~$\to$~CO~+~$\frac12$O$_2$ $\Delta H_f$=2.93~eV which is the ideal cost of 
producing one CO molecule. That is, $\epsilon$ which is larger than that value will 
lead to inevitable losses of energy not into the target process even at 100~\% 
conversion of CO$_2$ into CO. 

The CO$_2$ kinetics model applied for simulations here is incomplete in terms of chemistry -  
it does not include the secondary processes with O-atoms and reverse 
reactions. Also, interaction of CO$_2$ with reaction products 
CO, O$_2$, O is not taken into account. 
Therefore, the calculated $\chi$ is not expected to be 
a quantitatively correct prediction, but rather an indication that 
the process has been started. At the same time, the model can be considered 
as valid in the vicinity of the threshold values of $Q/n^2_0$ there the total 
CO$_2$ conversion is yet low and the influence of products on the primary 
dissociation process is expected to be small.
That is, the determination of the $Q/n^2_0$ threshold 
above which efficient dissociation from high vibrational states can start 
is the main result of the simulations, not the absolute 
values of $\chi$ at higher $Q/n^2_0$. 

All the simulations presented in figure~\ref{Qn2:Fig} are technically performed for the 
nominal pressure $p$=100~mbar. To confirm that, as suggested by~(\ref{concentration:master:eq}), 
there is no explicit pressure dependence the calculations 
were repeated with different values of $p$. 
The relative differences of $\chi$ calculated with $p$=50~mbar and 
$p$=200~mbar and those at $p$=100~mbar for the same values of $Q/n^2_0$
(only $\chi>$0.1~\% are taken into account) are found to be always $<$0.7~\%. 

The solid lines in figures~\ref{Qn2:Fig} present the solutions obtained with the reference model, 
and the dashed lines are the results of the simulations where the model uncertainties are evaluated. 
The issue that the SSH theory can grossly overestimate transition probabilities has 
been already discussed in section~\ref{shock:wave:S}. 
The potential impact of that shortcoming of the SSH scaling is estimated by reducing 
the imposed upped boundary of the SSH probabilities $p^{max}_{SSH}$ from the reference value 1 to 
0.1. The resulting effect is visible although relatively small for 
$T$=300~K, but tend to increase at higher $T$.  
Next, because of too large scattering of experimental data 
the pure theoretical rate coefficients are used for 
vibrational-vibrational transitions between symmetric modes - they are calculated 
by the Herzfeld theory~\cite{Herzfeld1967}. 
Experience shows that this theory is especially bad in predicting the absolute values 
of transition probabilities. Therefore, the consequences of reducing or increasing 
those coefficients had to be evaluated. This is done in the same way as in~\cite{KotovPSST2021}: 
the rate coefficients of the all corresponding transitions are multiplied by a factor $f^s_{VV}$. 
Its reference value is 1, $f^s_{VV}$=10$^{-2}$, 10$^{-1}$, 10  are tested. 
One can see, figure~\ref{Qn2:Fig}, a relatively large modification of the solution, 
especially when $f^s_{VV}$ is increased. At the same time, one can also say that both the 
variations of $p^{max}_{SSH}$ and $f^s_{VV}$ - although they have substantial impact on the 
steepness of the $\chi$ increase with increased $Q/n^2_0$ - have a very limited effect on the threshold 
value of this parameter. 

Finally, the effect of changing the electron temperature $T_e$ was investigated.  
In the reference model ($T_e$=1~eV) 69..81~\% of $Q$ is deposited into asymmetric modes. 
Two model runs have been added with $T_e$=0.5~eV and 2~eV. 
In the first case the fraction of $Q$ which goes into asymmetric modes is
slightly increased up to 80..92~\%, in the second case it is reduced down to 57..70~\%. 
As one can see in figure~\ref{Qn2:Fig} those variations have negligible impact on the solution.

\section{Approximate calculation of the process threshold $\left(Q/n^2_0\right)_{crit}$}

\label{approximate:threshold:S}

The critical value of the parameter $Q/n^2_0$ above which the non-equilibrium vibrational 
dissociation may be triggered can be calculated approximately on the basis of the 
simple considerations which will be put forward below. 
In the most general terms the balance of vibrational energy of the CO$_2$ molecules is   
written as:
\begin{equation}
\frac{ d \left( n E_{vibr} \right) }{d t}  =
Q - Q_{VT} - Q_{diss}
\label{evibr:general:eq}
\end{equation}
where $n$ is the molecule number density, $E_{vibr}$ is the average vibrational energy per molecule, 
$Q$ is the power deposited into vibrational modes, $Q_{VT}$ is the rate of the energy losses into 
translational-rotational modes, and $Q_{diss}$ is the rate of vibrational energy losses into 
chemical transformations. For the dissociation from the upper vibrational states to be significant 
their population has to be high enough, that is, $E_{vibr}$ has to be increased up to a certain high value. 
For this to happen at low translational-rotational temperature $T$ 
the right hand side of~(\ref{evibr:general:eq}) has to be larger than zero. 

In the near threshold regimes where $Q_{diss} < Q_{VT}$ the last term can be neglected,  
and the term $Q_{VT}$ can be written in the form suggested by equation~(\ref{evibr:relaxation:eq}):
\begin{equation}
 Q_{VT}\left(T, E_{vibr}\right)= R_{VT}\left(T\right) n_M n \left[ E_{vibr} - E^{eq}_{vibr}\left(T\right) \right]
 \label{Q:vt:empiric:eq}
\end{equation}
Where $n_M$ is the number density of the collision partners M in the VT-transfer process
~(\ref{bending:modes:transition:eq}) - here they are CO$_2$ molecules too,  
$R_{VT}\left(T\right)$ is the empiric energy relaxation rate coefficient 
obtained from the experimental characteristic times $\tau$, see section~\ref{shock:wave:S}.
In practice the coefficient $R_{VT}\left(T\right)$ 
can be calculated backward from the rate coefficients $R_{10}\left(T\right)$ 
of process~(\ref{bending:modes:transition10:eq}) published in the literature:
\begin{equation}
R_{VT}\left(T\right) = R_{10}\left(T\right) \left( 1 - e^{-\frac{\hbar\omega_2}{T}} \right)
\label{r_vt:backward:eq}
\end{equation}
as discussed in section~\ref{shock:wave:S}. 

\begin{figure}
\includegraphics[width=7cm]{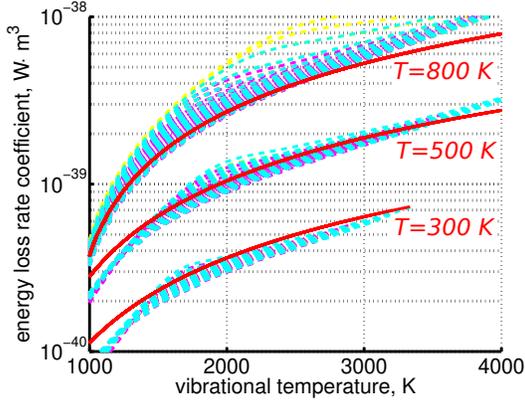}
\caption{Rate of vibrational energy losses $Q_{VT}$, see~(\ref{evibr:general:eq}), divided by $n_M n$.
         Solid line is the result of the approximate formula~(\ref{Q:vt:empiric:eq}), 
         dashed lines are $Q_{VT}/\left(n_M n\right)$ taken from numerical simulations of 
         section~\ref{model:runs:s}. Only modelling runs with the process rate
         $\chi<$12~\% are plotted}
\label{Q:VT:Fig} 
\end{figure}

Equation~(\ref{Q:vt:empiric:eq}) describes vibrational relaxation in gas dynamic experiments 
and, strictly speaking, must not be valid for the gas discharge conditions. 
In order to verify to which extent~(\ref{Q:vt:empiric:eq}) is applicable in that latter case too  
the magnitude of the term $Q_{VT}$ as it appears in numerical simulations 
of section~\ref{model:runs:s} 
is compared directly with that calculated by~(\ref{Q:vt:empiric:eq}). 
Examples of this comparison are shown in figure~\ref{Q:VT:Fig}. 
Instead of  $E_{vibr}$ the $Q_{VT}$ are plotted as functions of the more illustrative 
quantity $T_{vibr}$.
To translate the simulated $E_{vibr}$ into $T_{vibr}$ 
equations (A.2) and (A.3) from~\cite{KotovPSST2021} are used, 
$E_{vibr}$ in~(\ref{Q:vt:empiric:eq}) 
is calculated as a function of $T_{vibr}$ by applying~(\ref{evibr:CO2:harmonic:eq}).  
The $R_{VT}\left(T\right)$ in~(\ref{Q:vt:empiric:eq}) 
is calculated by using~(\ref{r_vt:backward:eq}), with $R_{10}$ 
taken from~\cite{BlauerNickerson1973}. To remind, the 2-modes model~\cite{KotovPSST2021}  
applied for the simulations 
is adjusted to match the experimental rates of vibrational relaxation  
which correspond to exactly this choice of  $R_{VT}\left(T\right)$,
see section~\ref{shock:wave:S}.

One can see in figure~\ref{Q:VT:Fig} that the agreement is relatively good.
This statement is quantified by examining  the ratio of 
$Q_{VT}\left(T_{vibr}\right)$ taken from the simulations
and $Q_{VT}\left(T_{vibr}\right)$ obtained with (\ref{Q:vt:empiric:eq}).  
For $T$=300..600~K this ratio always lies between 0.4 and 1.5. 
These maximum and minimum are taken over the all modelling runs of section~\ref{model:runs:s}  
including those with varied $T_e$, $p^{max}_{SSH}$ and $f^s_{VV}$ 
(only points with $T_{vibr}>T$+100~K are considered). 
For higher $T$ the upper boundary (maximum) of the ratio above is increased: up to 2.4
for $T$=800~K and up to 3.4 for $T$=1000~K. Nonetheless, even in those cases~(\ref{Q:vt:empiric:eq})
can always be taken as a good order of magnitude estimate.
This is an expected result since vibrational distribution function of the lower 
vibrational levels in the gas discharge simulations was found to be close to 
Boltzmann, see an example in~\cite{KotovPSST2021}. The populations of the upper 
levels near the dissociation limit deviate strongly from the Boltzmann distribution. 
However, since their populations are relatively small 
the high energy tail they form has no overwhelming impact on the total energy exchange rate.

At constant $T$ the term~(\ref{Q:vt:empiric:eq}) monotonically increases with increased $E_{vibr}$  
so that with $Q_{diss}$=0 the solution of~(\ref{evibr:general:eq}) would  
finally arrive at some required stationary value $E^*_{vibr}$ determined by the equality
$Q = Q_{VT}\left(T, E^*_{vibr}\right)$. 
Since only the very beginning of the conversion process is considered 
it can be taken that $n=n_M=n_0$, where  $n_0$ is the initial value before 
the process starts. 
Then the threshold (critical) value of $Q/n^2_0$ which corresponds to certain 
$E^*_{vibr}$ is readily found as:
\begin{equation}
\left( \frac{Q}{n^2_0} \right)_{crit} = R_{VT}\left(T\right) \left[ E^*_{vibr} - E^{eq}_{vibr}\left(T\right) \right]
\label{Qn2:crit:eq}
\end{equation}
The criterion~(\ref{Qn2:crit:eq}) has a simple physics meaning. 
For the vibrationally non-equilibrium process to start the 
rate of producing the vibrationally excited states has to be 
at least as fast as the rate with which vibrational energy is transferred into translational-rotational modes. 
 
The critical $E^*_{vibr}$  (or $T^*_{vibr}$) which enters~(\ref{Qn2:crit:eq})   
can be determined from the condition that at this $T^*_{vibr}$ 
the dissociation term $Q_{diss}$ starts to be comparable with $Q_{VT}$, e.g.
by assuming $Q_{diss}\left(T,T^*_{vibr}\right)=0.1Q_{VT}\left(T,T^*_{vibr}\right)$. 
Unlike $Q_{VT}$ there is no simple way of calculating $Q_{diss}\left(T,T^*_{vibr}\right)$, 
in particular because the dissociation proceeds from the 
non-Boltzmann high energy tail of vibrational distribution. 
Therefore, the determination of $T^*_{vibr}$ shall rely on the results of 
numerical simulations. The model uncertainties lead to a relatively 
large scattering of that quantity, but, as it will be seen below,  
this scattering has only minor impact on the final result. 

\begin{figure}
\subfloat[$T$=300~K]{
 \includegraphics[width=7cm]{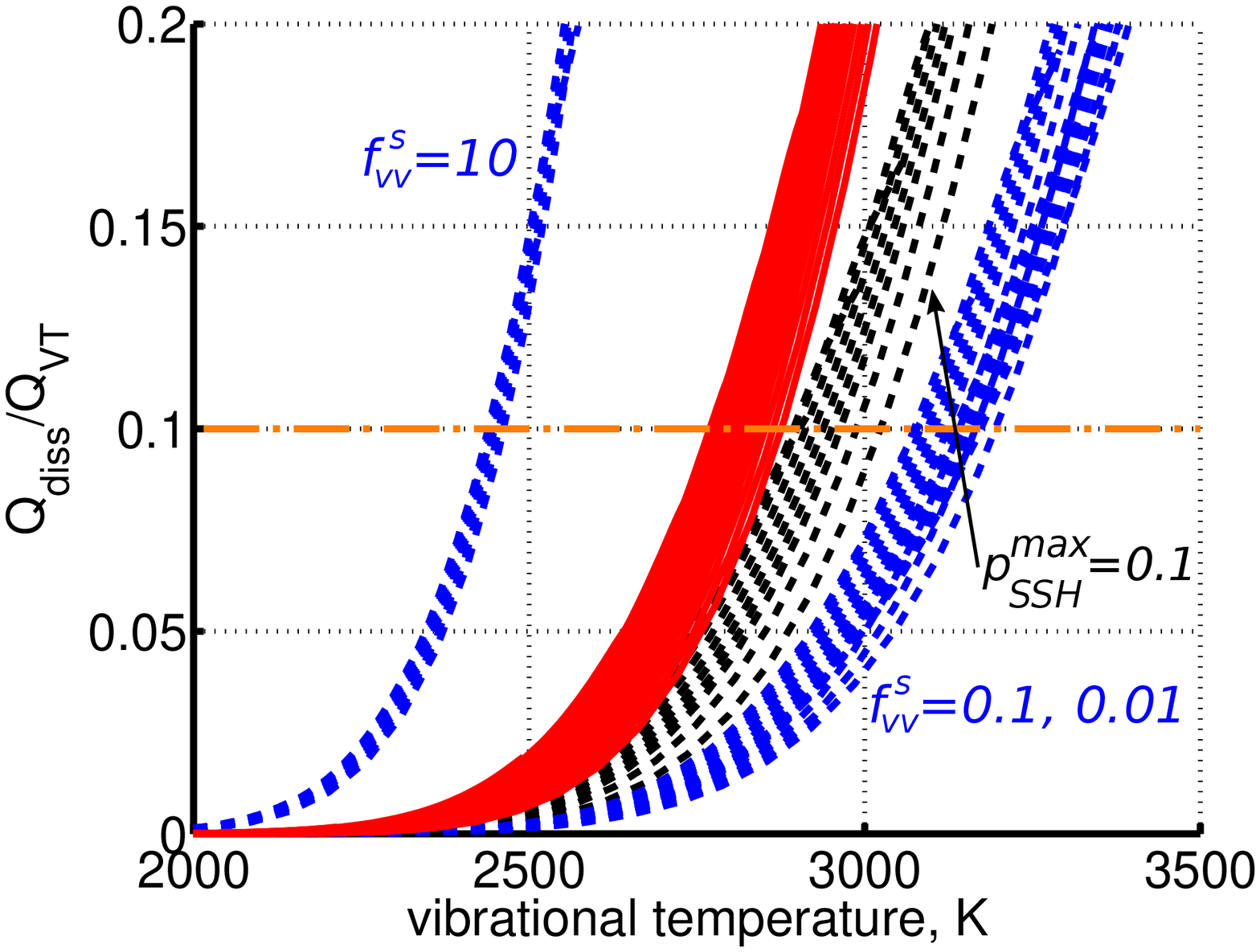} }
\subfloat[$T$=500~K]{
 \includegraphics[width=7cm]{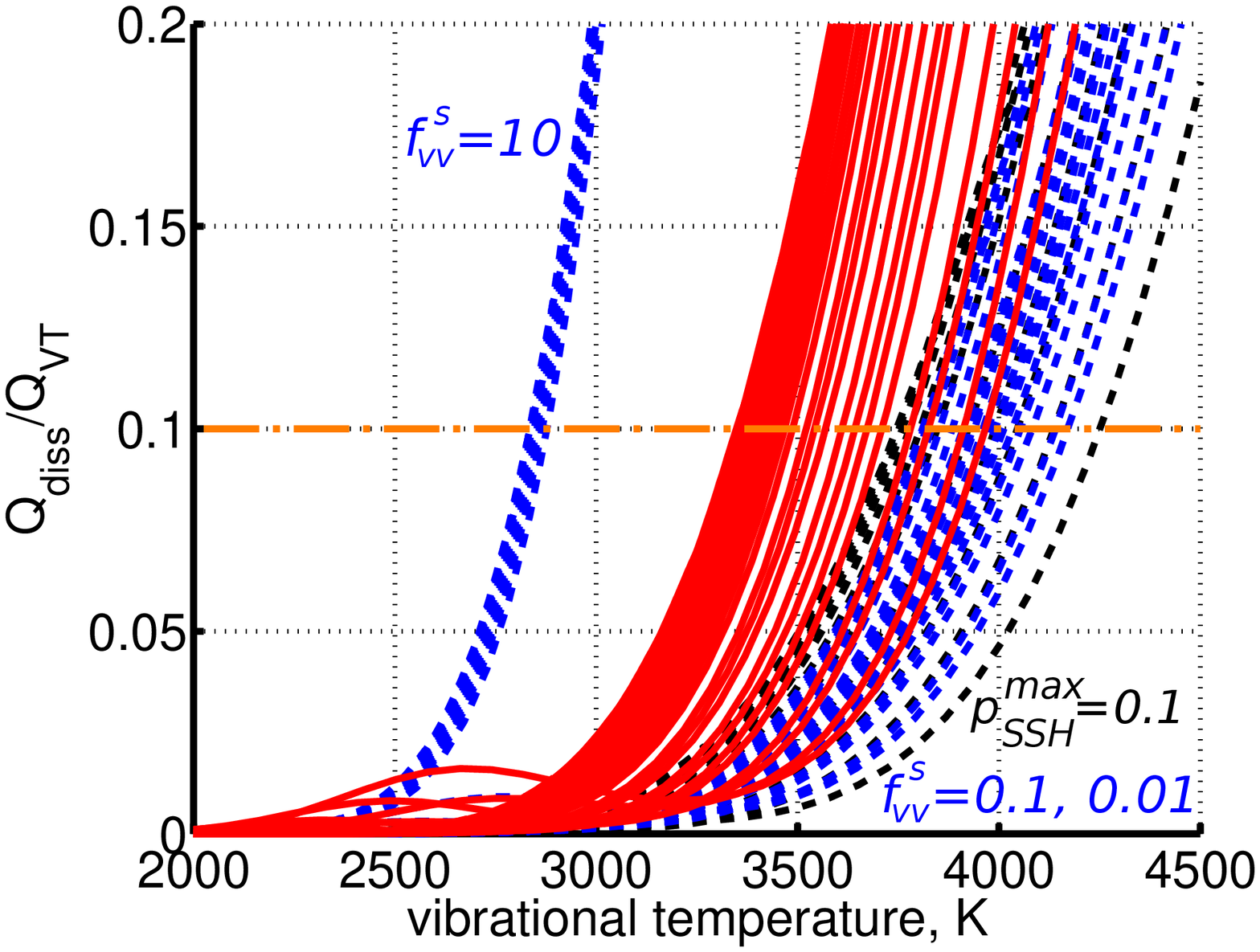} } \\
\subfloat[$T$=800~K]{
 \includegraphics[width=7cm]{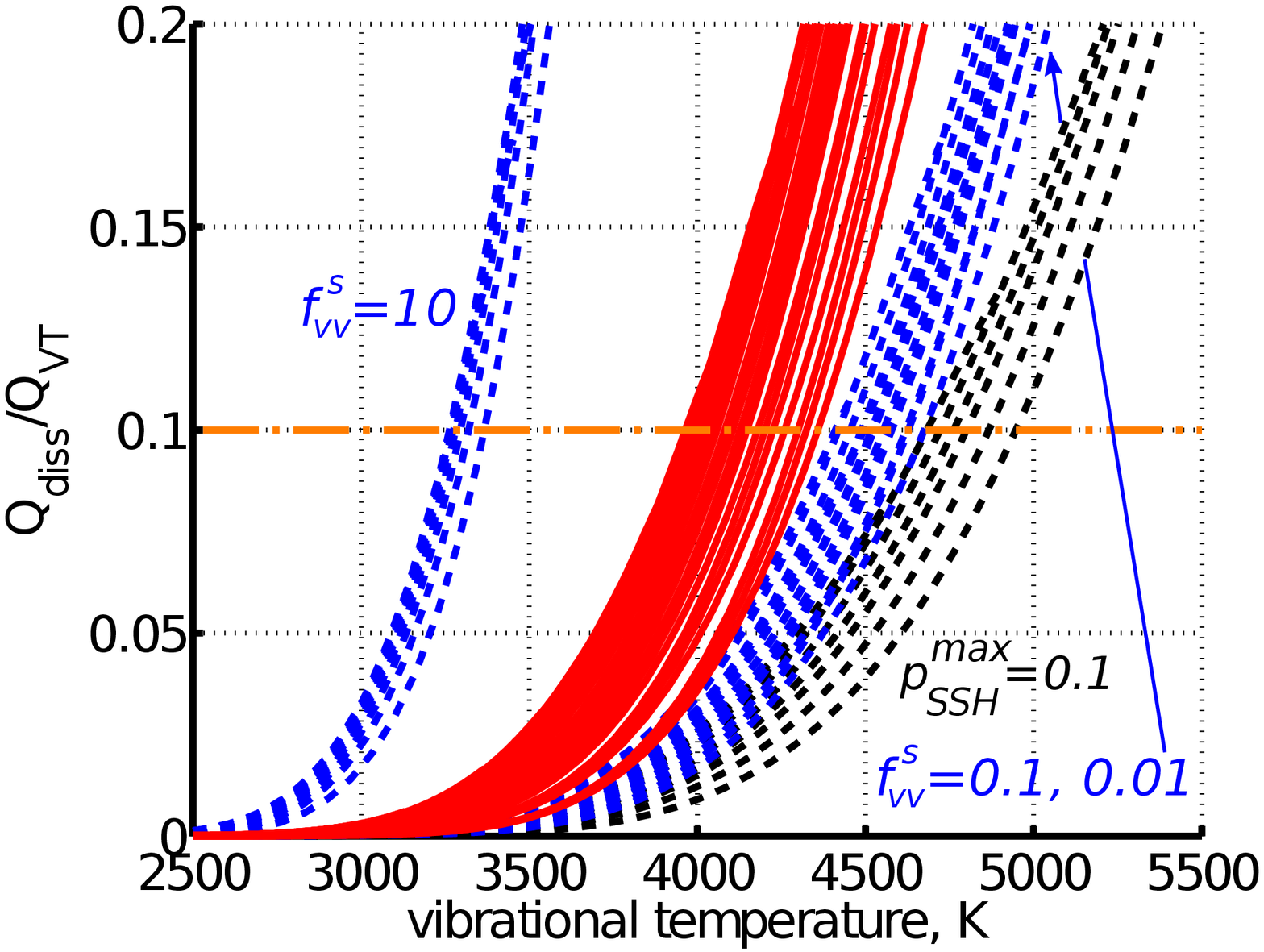} }
\caption{Ratio of the $Q_{diss}$ and $Q_{VT}$ terms, equation~(\ref{evibr:general:eq}), 
         as function of $T_{vibr}$ in numerical simulations of section~\ref{model:runs:s}.
         Solid lines correspond to the reference model with different values of $T_e$. 
         Dashed lines are the cases with  $p^{max}_{SSH}$=0.1 and
         $f^s_{VV}$=10$^{-2}$, 10$^{-1}$, 10. 
         Only modelling runs with the process rate $\chi<$25~\% are plotted}
\label{Q:diss:Fig} 
\end{figure}

\begin{table}
 \caption{Semi-empiric estimate of the critical value of  $Q/n^2_0$, 
 see section~\ref{approximate:threshold:S} }
 \begin{tabular}{l|l|l}
  \hline
  $T$, K & $T^*_{vibr}$, K & $\left(Q/n^2_0\right)_{crit}$, W$\cdot$m$^3$ \\
  \hline
   300 & 2500..3200 & (5..7)$\cdot$10$^{-40}$  \\
   500 & 2900..4300 & (2..3)$\cdot$10$^{-39}$  \\
   800 & 3300..5000 & (6..11)$\cdot$10$^{-39}$ \\
  \hline
 \end{tabular}
 \label{Qcrit:table}
 \end{table}

The ratio $Q_{diss}\left(T_{vibr}\right)/Q_{VT}\left(T_{vibr}\right)$ 
obtained in the simulations of section~\ref{model:runs:s}
is plotted for selected $T$ in figure~\ref{Q:diss:Fig}.  
The maximum and minimum of $T_{vibr}$ which correspond to 
$Q_{diss}\left(T_{vibr}\right)/Q_{VT}\left(T_{vibr}\right)\approx0.1$ 
for each $T$ are quoted in table~\ref{Qcrit:table}.
The numbers are rounded off because no high accuracy is required here.  
The corresponding values of $\left( Q/n^2_0 \right)_{crit}$ 
are obtained by applying equation~(\ref{Qn2:crit:eq})
with $E^*_{vibr}\left(T^*_{vibr}\right)$ calculated using~(\ref{evibr:CO2:harmonic:eq}).
One can see that the variation of $\left( Q/n^2_0 \right)_{crit}$ caused by the uncertainty 
of $T^*_{vibr}$ is not large. 
The increase of this parameter with increased $T$ is mainly due to increase of 
$R_{VT}\left(T\right)$ rather than the term in $[...]$ in~(\ref{Qn2:crit:eq}). 

The range of $\left( Q/n^2_0 \right)_{crit}$ from table~\ref{Qcrit:table}
is shown as shaded rectangles in figures~\ref{Qn2:Fig}.
The agreement with the process threshold as it appears in the results of the simulation runs is very good.
That is, equation~(\ref{Qn2:crit:eq}) together with the approximate values of $T^*_{vibr}$ 
provided in table~\ref{Qcrit:table} can be used as a practical criterion 
of activation of the non-equilibrium vibrational dissociation of CO$_2$. 
The accuracy of this approximate formula in determining the process threshold is not worse 
than the accuracy provided by a complicated vibrational kinetics model. 

\section{Summary and outlook}

In the present paper activation of the non-equilibrium vibrational dissociation of CO$_2$ 
at low translational-rotational temperatures $T$ in microwave gas discharges 
is investigated by means of computational and semi-empiric models.
A series of simulations has been performed with the 2-modes state-to-state model~\cite{KotovPSST2021}. 
The model has been benchmarked and adjusted against vibrational relaxation data 
from the shock tube and acoustic experiments. 
The governing parameter $Q/n^2_0$ has been introduced, 
where $Q$ is the volumetric specific power coupled in plasma, and $n_0$ is the initial number density of CO$_2$. 
The simulations indicate a rapid increase of the rate of the primary dissociation process 
CO$_2$~+~M~$\to$~CO~+~O~+~M with increased $Q/n^2_0$ when the governing parameter exceeds some critical (threshold) value. 
The solutions are found to be sensitive with respect to variation of the uncertain model parameters. 
However, the position of the process threshold as it appears in the simulation results is relatively stable. 
A simple analytic estimate of the critical value  $\left( Q/n^2_0 \right)_{crit}$ 
based on the experimental rates of vibrational relaxation has been proposed which 
agrees well with the modelling results. 

This estimate provides $\left( Q/n^2_0 \right)_{crit}$=6$\cdot$10$^{-40}$~W$\cdot$m$^3$ 
(averaged) at $T$=300~K. The numerical experiments suggest that within variation of the modelling result no 
non-equilibrium vibrational process is possible when  $Q/n^2_0<0.5\left( Q/n^2_0 \right)_{crit}$.
Efficient conversion of CO$_2$ into CO by the non-equilibrium process can be expected 
when $Q/n^2_0$ is larger than $\left( Q/n^2_0 \right)_{crit}$ withing an order of magnitude, see figure~\ref{Qn2:a:Fig}. 
For completeness it is worth to remind that in case it is applied 
e.g. for planning of experiments the 
$Q/n^2_0$ criterion is to be always supplemented by the straightforward Specific Energy Input (SEI) per molecule consideration. 
In experiments this quantity is typically controlled by the flow rate. 
SEI has to be at least larger than the energy per CO$_2$ molecule which corresponds to the target vibrational 
temperature $T_{vibr}$: e.g. for $T_{vibr}$=3000~K this is 0.8~eV. 
At the same time, SEI can not be allowed to be much larger than the net enthalpy change 
of the transition CO$_2$~$\to$~CO~+~$\frac12$O$_2$ $\Delta H_f$=2.93~eV 
because all the energy in excess of $\Delta H_f$ can not go into the target reaction. 

Formally, it may appear that the parameter  $\left( Q/n^2_0 \right)_{crit}$ can be easily increased 
by decreasing $n_0$ - that is, by reducing the gas pressure. 
In reality the pressure is restricted from below by at least two factors 
which were not taken into account in the present work.
The models applied here always implicitly assume optically thick gas. 
This assumption can be violated at very low pressures.
In that case a one more channel of vibrational energy losses via spontaneous emission will appear
in addition to collisional vibrational relaxation.
Second, at low pressures the shape of the energy distribution of electrons may shift towards 
higher energies,  and the power coupled in plasma will not go predominantly into 
vibrational excitations. The kinetic energy of electrons will be rather spent 
on other kinds of processes, such as direct electron impact dissociation. 
Proper account of the radiation transport and electron kinetics are two separate  
topics which were left completely out of the scope of the present work.
Models which couple self-consistently the non-maxwellian electron kinetics and the 
CO$_2$ vibrational kinetics and chemistry do exist~\cite{Capitelli2017,Pietanza2020}.
Their main weakness is that the 
cross sections of the electron processes with vibrational states of CO$_2$ are not known 
to a high degree of accuracy. 

The state-to-state model of the present paper is applicable only 
near the $Q/n^2_0$ threshold there the rate of conversion via the primary 
process CO$_2$~+~M~$\to$~CO~+~O~+~M is not yet too high. 
In order to further extend its applicability the secondary reactions with oxygen, 
reverse processes, reaction products and their influence on vibrational kinetics 
must be added. It can be expected that such extensions will not significantly affect the 
position of $\left( Q/n^2_0 \right)_{crit}$, but they might have a large impact on 
the steepness of the conversion rate growth with increased $Q/n^2_0$. 

Adding CO to the model can be seen as a pure technical matter since all the data
and models for calculation of the transition rate coefficients are readily available.
Adding the oxygen chemistry, on the opposite, can face large difficulties 
because of the exchange reaction CO$_2$~+~O~$\to$~CO~+~O$_2$. 
Too little is known about the influence of vibrational non-equilibrium on that process.
To which exactly extent  vibrational excitation of CO$_2$  speeds up this reaction? 
The process has a relatively low enthalpy change, where the vibrational energy of CO$_2$ will then go? 
Will it go into vibrational energy of the products or into their translational-rotational energy? 
Those questions have no definitive answers. 
The energy efficiency of the overall conversion process CO$_2$~$\to$~CO~+~$\frac12$O$_2$ 
can be larger than 50~\% only if the reaction CO$_2$~+~O~$\to$~CO~+~O$_2$ is faster 
than the 3-body recombination  O~+~O~+~M~$\to$~O$_2$~+~M, and if vibrational 
energy of CO$_2$ which takes part in the former goes predominantly into the products vibrational energy.
It is unlikely that all the unknowns will be resolved in the near future. 
Therefore, the most realistic next step in the state-to-state modelling studies of the CO$_2$ vibrational kinetics 
and chemistry is thought to be a model of a CO$_2$/CO mixture without oxygen. 
Particular questions this model could address are the 
effectiveness of vibrational coupling between CO$_2$ and CO, 
effective speed of vibrational relaxation of the mixture and 
the impact of CO on the primary dissociation process.

\section*{Data availability}

Link to the archive of the raw output of the modelling runs: 
\\ \url{https://doi.org/10.26165/JUELICH-DATA/IHO9NP} \\ 
The up-to-date source code of the numerical model (Fortran): \\
\url{https://jugit.fz-juelich.de/reacflow/co2-20} 

\appendix

\section{Derivation of vibrational relaxation equation for CO$_2$}

\label{vibrational:relaxation:CO2:S}

Derivation of the equation which describes thermal excitation of vibrational energy in diatomic molecules  
can be found in~\cite{HerzfeldLitovitz1959}, Chapter 19. 
Here the derivation of~\cite{HerzfeldLitovitz1959} 
is modified for CO$_2$ which has 3 modes of oscillation $v_1$, $v_2$, $v_3$. 
The mode $v_2$ is double degenerate and has rotational momentum $l$: $v^l_2$. 
It is assumed that the molecules are linear (harmonic) oscillators and their 
vibrational energy $E_v$ can be expressed as:
$$
E_v= \hbar\omega_1 v_1 + \hbar\omega_2 v_2 + \hbar\omega_3 v_3 
$$
Further assumptions are that vibrational-translational (VT) transfer takes place only 
from the mode $v^l_2$ via process~(\ref{bending:modes:transition:eq}), 
and the number of vibrational states is infinite.  

Since in the linear oscillator approximation $E_v$ does not depend on 
$l$ all states with same $v_2$ can be gathered into a combined state. 
There are $v_2+1$ different values of $l$ possible for a fixed $v_2$. 
The normalized matrix elements $X$ of all possible VT-transitions 
from the mode $v^l_2$ are summarized in~\cite{KotovJPB2020}, see equation (27) there:
$$
X_{v^l_2 \to \left(v_2-1\right)^{l+1}} =v_2-l; \quad 
X_{v^l_2 \to \left(v_2-1\right)^{l-1}}=v_2+l
$$
$$
X_{v^l_2 \to \left(v_2+1\right)^{l+1}}=v_2+l+2; \quad
X_{ v^l_2 \to \left(v_2+1\right)^{l-1} }=v_2-l+2
$$
The averaged probability $p$ of transition between combined states $v_2$ and $v_2-1$ 
calculated by the first order perturbation theory (SSH) is proportional to:
\begin{equation}
p_{v_2\to v_2-1} \sim \frac{\sum_l \left( X_{v^l_2 \to \left(v_2-1\right)^{l+1}} + X_{v^l_2 \to \left(v_2-1\right)^{l-1}} \right) }{ \sum_l 1  } \sim v_2
\label{p1:eq}  
\end{equation}
The sum $\sum_l$ is taken over all values of $l$ allowed for the corresponding $v_2$. 
Same for the transition between combined states $v_2$ and $v_2+1$:
\begin{equation}
p_{v_2\to v_2+1} \sim \frac{\sum_l \left( X_{v^l_2 \to \left(v_2+1\right)^{l+1}} + X_{v^l_2 \to \left(v_2+1\right)^{l-1}} \right) }{ \sum_l 1  } \sim 
 v_2 + 2 
\label{p2:eq}  
\end{equation}
Some transitions which appear in the sums are not possible. For example the transition from $v_2^{v_2}$ to $\left(v_2-1\right)^{v_2+1}$ 
is not possible because this latter state does not exist. Such transitions must not be excluded explicitly from the sums because the formulas for $X$ 
automatically give zero in those cases. 

Taking into account the assumption that only VT-transitions between modes $v^l_2$, process~(\ref{bending:modes:transition:eq}), 
are possible the particle balance equations for the number densities $n$ of the combined vibrational states 
$\left(v_1,v_2,v_3\right)$ 
read as follows: 
$$
\frac{dn_{v_1v_2v_3}}{dt} = k_{v_2+1\to v_2} n_{v_1 v_2+1 v_3} +  k_{v_2-1\to v_2} n_{v_1 v_2-1 v_3} -
$$
\begin{equation}
-\left( k_{v_2\to v_2-1} + k_{v_2\to v_2+1} \right) n_{v_1 v_2 v_3}, \quad v_2>0
\label{master1}
\end{equation}
\begin{equation}
\frac{dn_{v_10v_3}}{dt} = k_{10} n_{v_11v_3} - k_{01} n_{v_10v_3}
\label{master0}
\end{equation}
Here $k$ are the transition rates: transition rate coefficients multiplied by the total density of particles M 
in the process~(\ref{bending:modes:transition:eq}).  
The coefficients $k_{10}$ and $k_{01}$ are connected by the detailed balance relation:
\begin{equation}
\frac{k_{01}}{k_{10}} = \frac{w_1}{w_0}\exp{\left(-\frac{E_1-E_0}{T} \right)} = 2\exp{\left(-\frac{\hbar\omega_2}{T} \right)} = 2K
\label{detailed:balance}
\end{equation}
Here $w_j$ is the degeneracy of the vibrational level $\left(0,j,0\right)$ which is equal to $j+1$, and $E_j$ is its vibrational energy. 

All the other rate coefficients in~(\ref{master1}),~(\ref{master0}), 
can be connected with $k_{10}$ using the expressions~(\ref{p1:eq}),~(\ref{p2:eq})
and~(\ref{detailed:balance}):
\begin{equation}
k_{v_2\to v_2-1} = v_2 k_{10}, \quad k_{v_2\to v_2+1} =  \left(v_2+2\right) K k_{10}
\label{rate:coeff1}
\end{equation}
\begin{equation}
k_{v_2+1\to v_2} = \left(v_2+1\right) k_{10}, \quad k_{v_2-1\to v_2} = \left(v_2+1\right) K k_{10}
\label{rate:coeff2}
\end{equation}
Equations~(\ref{rate:coeff1}),~(\ref{rate:coeff2}) are substituted into equations~(\ref{master1}), (\ref{master0}). 
For convenience of notation indexes $v_1$, $v_2$, $v_3$ are replaced by $i$, $j$, $k$ respectively. The resulting master equations read:
\begin{equation}
\frac{dn_{ijk}}{dt} = k_{10} \left(j+1\right) n_{i j+1 k} +  k_{10} K \left(j+1\right) n_{i j-1 k} -
\label{reduced:master1}
\end{equation}
$$
- k_{10} j n_{ijk} - k_{10}K\left(j+2\right) n_{ijk}, \quad j>0
$$
\begin{equation}
\frac{dn_{i0k}}{dt} = k_{10} n_{i1k} - 2 K k_{10} n_{i0k}
\label{reduced:master0}
\end{equation}

To obtain the energy relaxation equation the~(\ref{reduced:master1}) and~(\ref{reduced:master0}) are multiplied by:
$$
E_v = \hbar\omega_1 i +  \hbar\omega_2 j + \hbar\omega_3 k = E_{ik} +   \hbar\omega_2 j
$$
and then their sum over indexes $i$, $j$, $k$ is calculated. This sum reads:
\begin{equation}
\frac{d \left[ \sum^\infty_{i,j,k=0} \left( E_{ik} +   \hbar\omega_2 j \right)  n_{ijk} \right] }{dt} = 
\label{evibr:equation:start}
\end{equation}
$$
= k_{10} \hbar\omega_2 \sum^\infty_{i,k=0} \sum^\infty_{j=1} \left\{ 
j\left(j+1\right) n_{i j+1 k} - j^2 n_{ijk} +  K j\left(j+1\right) n_{i j-1 k} 
 - Kj\left(j+2\right) n_{ijk} \right\} + 
$$
$$
+ k_{10} E_{ik} \sum^\infty_{i,k=0}\left[ \sum^\infty_{j=1} \left\{ 
\left(j+1\right) n_{i j+1 k} - j n_{ijk} +  K \left(j+1\right) n_{i j-1 k} 
 - K\left(j+2\right) n_{ijk} \right\} + n_{i1k} - 2 K n_{i0k} \right]
$$
The second term in~(\ref{evibr:equation:start}) equals zero. Indeed, combining the  terms
without $K$ in the innermost sum yields:
$$
\sum^\infty_{j=1}\left[ \left(j+1\right) n_{i j+1 k} - j n_{ijk} \right] +  n_{i1k} = 
\left| j'=j+1 \right| =
$$
$$
= \sum^\infty_{j'=2} j' n_{i j' k} +  n_{i1k} - \sum^\infty_{j=1} j n_{ijk}   = 
 \sum^\infty_{j=1} j n_{ijk} - \sum^\infty_{j=1} j n_{ijk} = 0 
$$
Same for the terms with factor $K$: 
$$
\sum^\infty_{j=1} \left[ \left(j+1\right) n_{i j-1 k} - \left(j+2\right) n_{ijk} \right]  - 2 n_{i0k} 
= \left|j'=j-1,\; j=j'+1 \right| =
$$
$$
= \sum^\infty_{j'=0} \left(j'+2\right) n_{i j' k} -  \sum^\infty_{j=1}\left(j+2\right) n_{ijk} - 2 n_{i0k}
= \sum^\infty_{j=0} \left(j+2\right) n_{i j k} -  \sum^\infty_{j=0}\left(j+2\right) n_{ijk} = 0
$$
The sum:
$$
 \sum^\infty_{i,j,k=0} \left( E_{ik} +   \hbar\omega_2 j \right)  n_{ijk} = E_{vibr} n,\quad n = \sum^\infty_{i,k,j=0} n_{i j k}
$$
is the total specific (per unit volume) vibrational energy; $E_{vibr}$ is the specific vibrational energy per molecule, $n$ is the total number density 
of molecules. The whole equation~(\ref{evibr:equation:start}) is reduced to the following form:
$$
\frac{d \left( n E_{vibr} \right) }{dt} = 
  k_{10} \hbar\omega_2 \sum^\infty_{i,k=0} \sum^\infty_{j=1} \left[ j\left(j+1\right) n_{i j+1 k} - j^2 n_{ijk} \right] +
$$
\begin{equation}
 +  k_{10} \hbar\omega_2 K \sum^\infty_{i,k=0} \sum^\infty_{j=1} \left[ j\left(j+1\right) n_{i j-1 k}  - j\left(j+2\right) n_{ijk} \right]
 \label{evibr:equation:interm}
\end{equation}
Calculation of the first sum of~(\ref{evibr:equation:interm}):
$$
\sum^\infty_{j=1} \left[ j\left(j+1\right) n_{i j+1 k} - j^2 n_{ijk} \right] =  \left|j'=j+1,\; j=j'-1 \right| =
$$
$$
= \sum^\infty_{j'=1} \left(j'-1\right) j' n_{i j' k} -  \sum^\infty_{j=1} j^2 n_{ijk} = 
 \sum^\infty_{j=1} \left[ \left(j-1\right) j - j^2 \right] n_{i j k} = -\sum^\infty_{j=0} j n_{i j k}
$$
Calculation of the second sum of~(\ref{evibr:equation:interm}):
$$
\sum^\infty_{j=1} \left[ j\left(j+1\right) n_{i j-1 k}  - j\left(j+2\right) n_{ijk}\right] = 
\left| j'=j-1,\; j=j'+1 \right| = \sum^\infty_{j'=0} \left(j'+1\right) \left(j'+2\right) n_{i j' k} -
$$
$$
 - \sum^\infty_{j=0} j\left(j+2\right) n_{ijk} =
   \sum^\infty_{j=0}\left[ \left(j+1\right) \left(j+2\right)  -  j\left(j+2\right) \right] n_{i j k} = \sum^\infty_{j=0} \left(j+2\right) n_{i j k}
$$
Substituting the calculated sums back into~(\ref{evibr:equation:interm}) yields:
$$
\frac{d \left( n E_{vibr} \right) }{dt} =  k_{10} \hbar\omega_2
\left[   K  \sum^\infty_{i,k,j=0} \left(j+2\right) n_{i j k}  -  \sum^\infty_{i,k,j=0} j n_{i j k} \right] =
$$
\begin{equation}
=  k_{10} n \left[ K E_{vibr-2} + 2 \hbar\omega_2 K - E_{vibr-2} \right]
\label{evibr:equation:semifinal}
\end{equation}
where:
$$
\sum^\infty_{i,k,j=0} \hbar\omega_2 j n_{i j k} =  n E_{vibr-2} 
$$
is the total vibrational energy stored in the mode $v^l_2$, $E_{vibr-2}$ is its specific value per molecule. 
Transforming the right hand side of~(\ref{evibr:equation:semifinal}) and substituting $K$, equation~(\ref{detailed:balance}), 
yields finally:
$$
\frac{d \left( n E_{vibr} \right) }{dt} =   k_{10}n\left[2 \hbar\omega_2  K - \left( 1- K \right) E_{vibr-2} \right] =
                                            k_{10}n \left(1-K\right) \left[ \frac{2  \hbar\omega_2  K}{1-K} - E_{vibr-2} \right] =
$$
$$
 = k_{10}n \left( 1 - e^{- \frac{\hbar\omega_2}{T}} \right)
  \left[ \frac{2  \hbar\omega_2 }{e^{\frac{\hbar\omega_2}{T}} - 1} - E_{vibr-2} \right]
$$ 
For $n=const$ and with $k_{10}$ replaced by $R_{10}\left(T\right)n$ this equation gives~(\ref{evibr:CO2:relaxation:eq}),~(\ref{evibr2:harmonic:eq}).


\end{document}